\documentclass[12pt]{article}
\usepackage{enumitem,multirow,multicol}
\usepackage{setspace}
\usepackage{amssymb,amsmath,amsthm}

\newtheorem{theorem}{Theorem}

\usepackage{mathtools} 

\usepackage{graphicx,xcolor}
\graphicspath{{img/}}
\usepackage{amsmath, bm}
\usepackage[scr=rsfs]{mathalpha}

\newtheorem{proposition}{Proposition}

\newtheorem{corollary1}{Corollary}

\newtheorem{remark1}{Remark}

\usepackage{sectsty}
\usepackage{epsfig}
\usepackage{rotate}
\usepackage{lscape}
\usepackage{setspace}
\usepackage{subcaption}
\usepackage{float}
\usepackage{mathtools}
\usepackage{booktabs}
\usepackage{natbib}

\renewcommand\thesection{\arabic{section}.}
\renewcommand{\thesubsection}{\thesection\arabic{subsection}}
\sectionfont{\centering\bf\sf\normalsize}
\subsectionfont{\sf\normalsize}

\def\bSig\mathbf{\Sigma}

\DeclareMathOperator*{\argmin}{argmin}
\newcommand{\bea}{\begin{eqnarray*}}
\newcommand{\eea}{\end{eqnarray*}}
\newcommand{\be}{\begin{eqnarray}}
\newcommand{\ee}{\end{eqnarray}}

\newcommand{\no}{\noindent}
\newcommand{\bc}{\begin{center}}
	\newcommand{\ec}{\end{center}}

\usepackage[figuresright]{rotating}

\begin{document}
	
\setcounter{footnote}{2}
\thispagestyle{empty}  \bc {\bf \sc \Large Latent Deformation Models for Multivariate Functional Data and Time Warping Separability}

\vspace{0.5in}

Cody Carroll$^1$,  Hans-Georg M\"uller$^2$\\
$^1$Department of Mathematics and Statistics, University of San Francicso \\
$^2$Department of Statistics, University of California, Davis \\

\ec \centerline{2023}

\vspace{0.4in} \thispagestyle{empty}
\bc{\bf \sf ABSTRACT} \ec \vspace{-.1in} \no 
\setstretch{1}
Multivariate functional data present theoretical and practical complications which are not found in  univariate functional data. One of these is a situation  where the component functions of multivariate functional data are positive and are subject to mutual time warping. That is, the component processes exhibit a common shape but are subject to systematic phase variation across their domains in addition to subject-specific time warping, where each subject has its own internal clock. This motivates a novel model for multivariate functional data that connects such mutual time warping to a latent deformation-based framework by exploiting a novel time warping separability assumption. This separability assumption  allows for meaningful interpretation and dimension reduction. The resulting Latent Deformation Model is shown to be  well suited to represent commonly encountered functional vector data.  The proposed approach combines a random amplitude factor for each component with population based registration across the components of a multivariate functional data vector and includes a latent population function, which corresponds to  a common underlying trajectory. We propose estimators for all components of the model, enabling implementation of the proposed  data-based representation for multivariate functional data and  downstream analyses such as  Fr\'echet regression.  Rates of convergence are established  when curves are fully observed or observed with measurement error. The usefulness of the model, interpretations, and practical aspects are illustrated in simulations and with application to multivariate human growth curves and  multivariate environmental pollution data.\\

\no {KEY WORDS:\quad Component processes, cross-component  registration, functional data analysis,  longitudinal studies, multivariate functional data,   time warping}.


\newpage
\pagenumbering{arabic} \setcounter{page}{1}

\newpage
\pagenumbering{arabic} \setcounter{page}{1} 

\setstretch{1.5}


\label{firstpage}


%
%

\section{Introduction}
\label{s:intro}
Functional data analysis (FDA) has found important applications  in many fields of research (e.g. biology, ecology, economics) and  has spawned considerable methodological work as a subfield of statistics \citep{rams:05, mull:16:3, ferr:06}. In particular, the analysis of  univariate functional data has driven the majority of developments in this area  such as functional principal component analysis \citep{klef:73}, regression \citep{card:99,yao:05}, and clustering \citep{chio:07,jacq:14}. 
In this paper we develop novel modeling approaches for  multivariate functional data, which  consist of samples of a finite dimensional vector whose elements are random functions \citep{chio:14, jacq:14} and have  been much less studied. 
Dimension reduction is a common approach, with many studies focusing on extending univariate functional principal components analysis to the multivariate case \citep{happ:18,mull:19:1} and decomposition into marginal component processes and their interactions \citep{chio:16}. 

Most  methodological work has focused on traditional amplitude variation-based models for dimension reduction, while phase variation-based methods for multivariate functional data have found attention more recently:  \cite{brun:14}  proposed a method for estimating multivariate structural means and \cite{park:17} introduced a model for clustering multivariate functional data in the presence of phase variation, while  \cite{carr:21} combined the notions of dimension reduction and phase variability through a multivariate version of the shape-invariant model \citep{knei:95}, in which component processes share a common latent structure that is time-shifted across components. However,  the assumption of a rigid shift-warping framework in this precursor work  imposes a major parametric constraint on the warping structure and often the class of models that only feature  simple shifts between the components is not rich enough for many real-world data. Our main contribution is a less-restrictive alternative approach,  in which time characterization of individual-specific temporal effects and component-specific effects is achieved through a fully non-parametric deformation-based model.  

A major motivation for this framework is  that in many contexts, the component functions of a multivariate data vector  may share a common structure that is subject to variation across modalities; the fundamental shape of growth curves is similar but not identical when studying  timing patterns across body parts, for instance. 
A reviewer suggested to alternatively align the components for each subject in a constrained way; we demonstrate in this paper  that an overall more compelling model is obtained 
by assuming a latent common curve is present at  the population level, which brings with it the benefits of dimension reduction and a principled and novel representation of mutually time-warped functional data.
 
The proposed  latent curve model introduces  a shared shape-based model along with  a  characterization of individual- and component-level variation 
and allows for  flexible and nuanced component effects. This ensures broad viability of the proposed approach and improved data fidelity when describing component-specific effects,  which inform the time dynamics of a larger system at work. To this end, we introduce a representation of multivariate functional data which uses tools from time warping \citep{marr:15} and template deformation modeling \citep{bigo:09, bigo:11}.

The organization of this paper is as follows. Section 2 discusses existing approaches for univariate curve registration and introduces the proposed Latent Deformation Model for component-warped multivariate functional data. We derive estimators of model components in Section 3 and illustrate the utility and performance of the proposed methodology through data analysis in Section 4. Asymptotic results are established in Section 5, and a discussion of goodness-of-fit issues and a simulation study are provided in the Appendix, which also contains auxiliary results and proofs.



\section{Curve Registration and The Latent Deformation Model}

The main idea of the Latent Deformation Model (LDM) that we introduce in this paper is to decompose multivariate phase variation into subject-specific and variable-specific warping components. When combined with a common, shape-defining template, these warping functions provide a lower-dimensional representation of the functional vector trajectories while characterizing the subject-level warping and population-wide patterns in the time dynamics across variables. In addition to the existence of a template function shared across subjects, the proposed LDM  includes a modeling assumption  that each subject has an ``internal clock,"  which is quantified through a subject-specific warping function. Similar assumptions have been previously explored in the cross-component registration paradigm of \cite{carr:21}, which however restricted the component-wise phase variation to simple parametric shift functions. A major contribution of this paper is to widen the class of potential component warps beyond rigid shifts to allow for more flexible warping functions, so as to better capture variation that  occurs non-uniformly across the time domain. 

Before introducing the detailed mathematical machinery of the model, a brief overview of the general idea is as follows. We first introduce  a flexible and separable component structure for warping functions, which are factorized into subject- and component-specific warpings  and then proceed to develop  estimates of  these factor warping functions.   The first step is to construct consistent estimates of the subject-specific warping functions which correspond to the internal clock of each subject. This is done by considering univariate warping problems for each functional variable separately and then averaging the resulting estimates of the component warping functions for each subject, resulting in a consistent estimate of the  subject-specific time warping function.  Eventually this then  leads to  consistent estimates of the underlying latent curve. Assuming  that  time-warped versions of this underlying latent  curve generate  the functional vector component-level distortions, in order to recover it, one component function is selected at random per subject, discarding the data from the other components, then aligning these curves across subjects. Once this consistent estimate of the underlying template curve has been obtained, consistent estimates of the  component-level distortion functions are recovered by solving a penalized cost minimization problem.   A schematic of the data generating mechanism of the LDM is provided in Figure 1. More detailed descriptions follow below. 

\begin{figure}[!p]
	\centering
	\vspace{-10mm}
	\includegraphics[width=\linewidth]{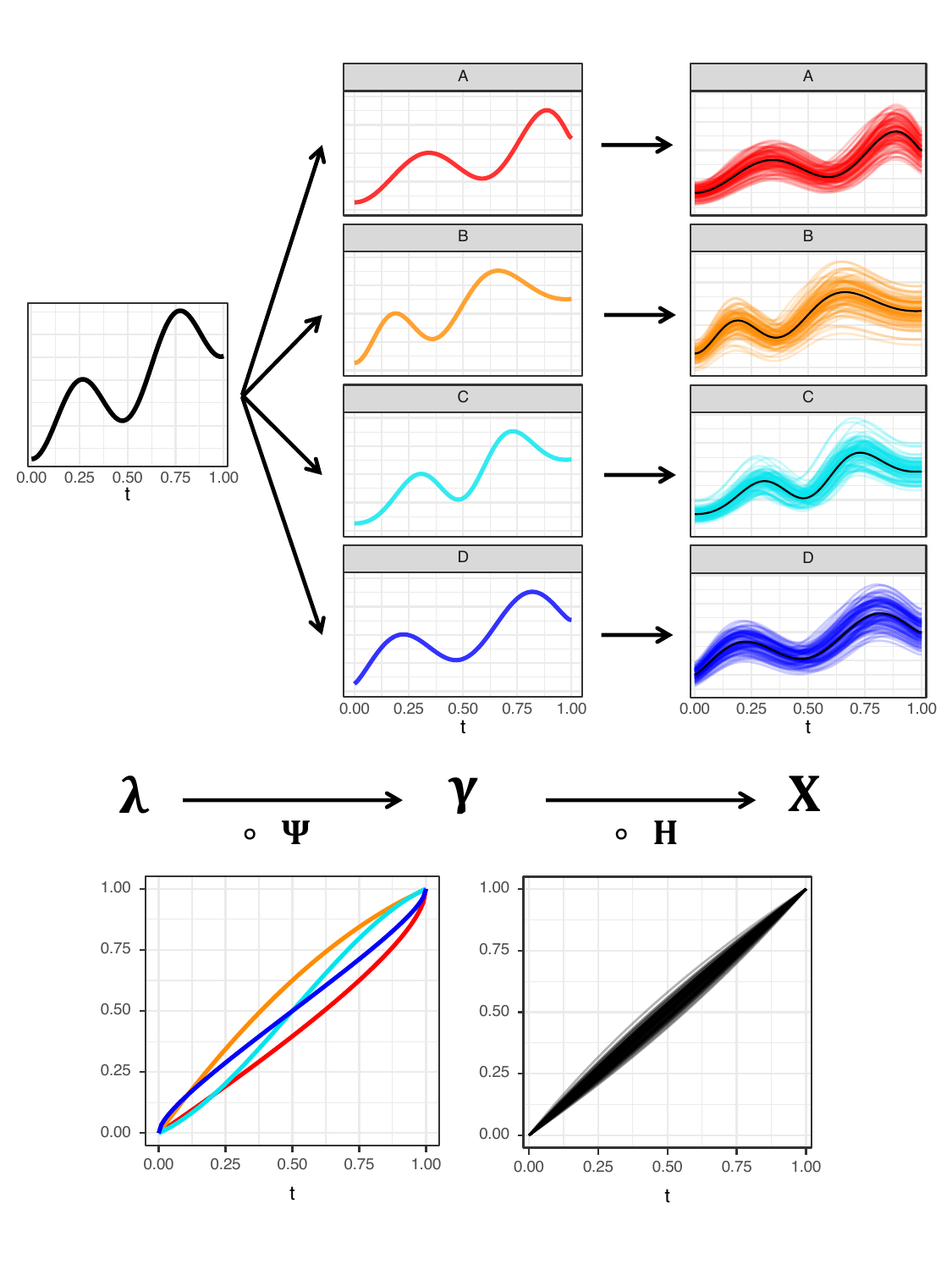}
	\caption{Schematic of the Latent Deformation Model, where $\lambda$  denotes the latent base curve (top-left), $\Psi$ denotes component deformations (bottom-left), $\gamma$ denotes component tempos (top-center), $H$ denotes random subject-wise time distortion functions (bottom-right), and $X$  denotes the observed multivariate curve data (top-right) resulting from the complete data generating mechanism. }
	
	\label{fig:figure1}
\end{figure}

\subsection{The Univariate Curve Registration Problem}

The classical univariate curve registration problem is characterized by the observation of a sample of curves $X_i(t),~ i = 1,\dots,n,$ observed on an interval $T$,  which are realizations of a fixed template $\xi(t)$ subject to variation in their time domains. This domain variation is characterized by the monotonic time-warping functions $h_i(t)$ which act as random homeomorphisms of $T$. A classical model for this scenario is 
\begin{equation}
X_i(t) = (\xi \circ h_i)(t), \quad \text{for all} \quad t\in \mathcal{T},\quad i=1,\dots,n.
\label{eq.1}
\end{equation}
The goal of curve registration is to estimate the distortions, $h_i$, which are typically considered nuisance effects, in order to account for them before proceeding with further analysis, e.g.,  estimation of $\xi$, functional principal component analysis, etc. 
A major branch of time-warping techniques is based on the idea of aligning processes to some reference curve which carries the main features in common across subjects. This reference curve is referred to as a template function and is employed by landmark-based registration methods \citep{knei:92, knei:95}, pairwise curve alignment \citep{mull:08:6} or the Procrustes approach \citep{rams:98}, among many others.  For a comprehensive review and additional references we refer to \cite{marr:15}.  


While the curve registration literature is varied and rich in methodology, no single method has prevailed as a silver bullet in all warping contexts. Indeed, the debate over desirable properties of existing and future registration techniques continues and a gold-standard remains elusive. With this in mind, we emphasize that our aim here is not to advocate for one alignment method over another, but rather extend the ideas available for univariate registration to a multivariate problem with a composite warping function with fixed and random effects. 
In practice, any suitable registration method may be employed in the estimation step of the proposed Latent Deformation Model (see Estimation).  

\subsection{A Unified Model for Multivariate Time Dynamics and Time Warping Separability}

Let $\{X_j\}_{j=1}^p$ denote a generic set of random functions with each component process $X_j$ in $L_2(\mathcal{T})$ for an interval $\mathcal{T} = [T_1,T_2], ~T_1,T_2\in\mathbb{R}$.  Suppose further that each component is positive-valued, i.e. $X_j(t)>0$ for all $t \in \mathcal{T}$, $j=1,\dots,p$; the assumption of positivity is made to make estimation of model components more straightforward and is certainly satisfied for applications to growth curves.  Without loss of generality we consider  the unit domain case $\mathcal{T} = [0,1]$. In the following,  Greek letters denote fixed, unknown population quantities, while Latin letters represent random, individual-specific quantities. 

The Latent Deformation Model (LDM) is motivated by situations where the functional forms of the component processes $X_j,~ j\in\{1,\dots,p\}$ (or any subset thereof) exhibit structural similarity, so that the information inherent in each component may be combined for overall improved model fitting and to estimate and analyze the mutual time warping structure. Denoting a random sample from a $p$-dimensional stochastic process by $\{\textbf{X}_i\}_{i=1}^n$, where $\bm{X}_i(t) = (X_{i1}(t),\dots,X_{ip}(t))^T$, we model this shared structure through a latent curve $\lambda$, which characterizes the component curves through the relation
\begin{equation}
X_{ij}(G^{-1}_{ij}(t)) = A_{ij} \lambda(t) , \quad  i = 1,\dots, n,~j = 1,\dots, p,
\label{eq.2}
\end{equation}
\noindent where $\lambda$ is a fixed function, and the random amplitude factors $A$ and random time distortion functions $G$ reflect differences in realized curves across components and individuals. Without loss of generality we assume $\underset{t \in \mathcal{T}}{\sup}|\lambda(t)|=||\lambda||_\infty = 1 $ since it is always possible to rescale the latent curve without changing the model by employing amplitude factors $\tilde{A}_{ij} := A_{ij} ||\lambda||_\infty$ and a standardized  curve $\tilde{\lambda}(t) = \lambda(t)/||\lambda||_\infty$. 

The distortion functions $G$ are elements of $\mathcal{W}$, the convex space of all smooth, strictly increasing functions with common endpoints, i.e., $\mathcal{W}:=\{g:\mathcal{T}\to \mathcal{T}~ |~g\in C^2(\mathcal{T}), ~ g(T_1)=T_1, ~ g(T_2)=T_2,~ g~\text{is a strictly increasing homeomorphism}\}$.  The elements of this space represent random homeomorphisms of the time domain and capture the presence of non-linear phase variation. We further assume that the distortion functions $G$ may be decomposed as follows, 
\begin{equation}
G_{ij}(t) = (\Psi_j \circ H_i) (t)  , \quad  i = 1,\dots, n,~ j = 1,\dots, p, 
\label{eq.3}
\end{equation}
\noindent where the deterministic functions $\Psi$ describe the component-based effects of time distortion and the random functions $H$ describe the subject-level phase variation. 

This decomposition is key to our approach.  A reviewer suggested to refer to it as a  separability assumption and indeed it is analogous to the well-known notion of separability of covariance in function-valued stochastic process modeling \citep{mull:17:4,lian:22} 
and we have  adopted this suggestion,  as it brings out a key aspect of the proposed LDM. 
As in the related covariance separability paradigm, time warping separability confers the advantages of better interpretability and dimension reduction over the more complex approaches that do not include this assumption.

Under the warping separability assumption, the time warping functions  $G_{ij}$ are decomposed into the warping maps  $\Psi_j$ that convey  the relative  time scale of the $j^{th}$ component and  the warping maps  $H_i$ that  quantify the internal clock of the $i^{th}$ subject. These warping maps can be viewed as deformations from standard clock time, $id(t)\equiv t$, to the system time of a given component or individual. As such we refer to the collection of functions $\varPsi = \{\Psi_j: j = 1,\dots, p\}$ as \textit{component-level deformation functions} and the collection of functions $\mathcal{H} = \{H_i: i = 1,\dots, n\}$ as \textit{subject-level deformation functions}.

The random subject level deformation functions $H_i$ obey some probability law on the convex space $\mathcal{W}$,  where we assume that this probability law is such that  $EH^{-1}_i$ exists and  that there is  no net distortion on average, i.e.,
$EH^{-1}_i(t) = t$ for $t \in \mathcal{T}$. This assumption has been  referred to as ``standardizing" the registration procedure \citep{knei:08}. It is a mild assumption, since were it the case that $EH^{-1}_i(t) = {h}_0^{-1}(t),~$ with $ {h}_0^{-1}\neq id $, then a standardized registration procedure is given by reparameterizing the warping functions as $\tilde{H}_i = h_0^{-1}\circ H_i$ so that $E\tilde{H}^{-1}_i(t) = E(H^{-1}_i\circ h_0)(t)= t$. Component deformation functions are also assumed to be standardized, but because they are deterministic and not random, the assumption becomes  $\frac{1}{p}\sum_{j=1}^p\Psi^{-1}_j (t) = t$ for $t \in \mathcal{T}$. Together these conditions imply $E(\frac{1}{p}\sum_{j=1}^p G^{-1}_{ij} (t)) = t$ so that there is no net distortion from the latent curve $\lambda$. 

Combining (\ref{eq.2}) and (\ref{eq.3}) yields the Latent Deformation Model (LDM)  for multivariate functional data, given by 
\begin{equation}
X_{ij}(t) = A_{ij} \left(\lambda \circ \Psi_j \circ H_i\right) (t)  , \quad i = 1,\dots, n, ~ j = 1,\dots, p. 
\label{eq.4}
\end{equation}
\noindent In practice, it may be useful to pose the model in an equivalent form, defining the component-warped versions of the latent curve as $\gamma_j = \lambda \circ \Psi_j$ so that
\begin{equation}
X_{ij}(t) = A_{ij} \left(\gamma_j \circ H_i\right) (t) , \quad i = 1,\dots, n, ~ j = 1,\dots, p. 
\label{eq.5}
\end{equation}
In this form, the curves $\gamma_j(t)$ convey the ``typical" time progression of the latent curve according to the $j^{th}$ component's system time, so we refer to this composition as the $j^{th}$ \textit{component tempo function}. The component tempo functions can be viewed as the synchronized processes for each component after accounting for random subject-level time distortions.



\subsection{Cross-Component Deformation Maps}
\label{sec:xct} 

\noindent\textit{Marginal Cross-Component Deformations}\\
To understand and quantify the relative timings between any pair of components, $j,k\in\{1,\dots,p\}$, it is useful to define their \textit{cross-component deformation} $T_{jk}$, which is the deformation that, when applied to the $j^{th}$ component, maps its tempo to that of the $k^{th}$ component,
\begin{equation}
T_{jk} = \Psi^{-1}_{j}\circ\Psi_{k}, 
\end{equation}
so that $\gamma_j(T_{jk}) = \lambda \circ \Psi_{j} \circ \Psi^{-1}_{j} \circ \Psi_{k}  = \lambda \circ \Psi_{k} = \gamma_k$. Because the component deformations $\Psi_{k}$ can be represented as distribution functions and are closed under composition, the cross-component deformation (XCD) may also be represented as a distribution function and is interpreted similarly to an ordinary component tempo. While the component tempo $\Psi_k$ expresses the $k^{th}$ component's timing patterns in terms of clock time, the cross-component deformation $T_{jk}$ expresses the same patterns relative to the tempo of the $j^{th}$ component. 

For example, consider a pair of component processes, Component A and Component B, for which Component A tends to lag behind the latent curve, while the Component B precedes it. An example of this can be seen in the red and orange curves, respectively, in Figure \ref{fig:figure1}. The corresponding red deformation, $\Psi_{A}$, falls below the diagonal and conveys the lagged tempo, while the orange deformation, $\Psi_{B}$ lies above the diagonal and expresses an accelerated system time. The deformation function $T_{AB}$ then sits above the diagonal and represents the time-acceleration needed to bring the red tempo in line with the orange component. 

\noindent\textit{Subject-Level Cross-Component Deformations}\\
While the marginal XCDs describe the general time relations between components on a population level, we may also be interested to see how an individual's component processes relate to one another. This perspective may be especially useful when trying to understand  intercomponent dynamics which are mediated by covariate effects. Conceptually it is straightforward to extend the notion of cross-component deformations to individuals by searching for the warping function $T^{(i)}_{jk}$ which brings the $i^{th}$ individual's $j^{th}$ component in line with the $k^{th}$. A natural definition under the LDM is then
\begin{equation}
T^{(i)}_{jk} = G^{-1}_{ij} \circ G_{ik},
\label{indiv_xct}
\end{equation}\noindent
since this choice gives $X_{ij}\circ T^{(i)}_{jk} \propto A_{ij} (\lambda \circ G_{ij} \circ G^{-1}_{ij} \circ G_{ik}) \propto (\lambda \circ G_{ik}) \propto X_{ik}$. In practice, this proportionality will become equality once random amplitude factors are dealt with during estimation.  Statistics based on the XCDs can be used in downstream analyses like hypothesis testing and regression. Several data illustrations are given in the applications of Section~\ref{sec:data}.
\vspace{-1cm}
\section{Model Estimation and Curve Reconstruction}
\label{sec:est}
\subsection{Internal Clock Estimation and Component-wise Alignment}

The proposed model estimation procedure relies on solving several univariate warping problems of type $(1)$. It is important to note that any of the warping methods described in Section 2 may be used for
practical implementation. In our implementation we choose  the pairwise alignment method of \cite{mull:08:6}, which provides an explicit representation of the warping functions and satisfies some properties required by our theory in order to derive convergence rates. This pairwise alignment  is  easily  implemented with  the R package \texttt{fdapace} \citep{carr:20fdapace}. For a  detailed discussion of the pairwise warping method we refer to the supplement. 

For the estimation of the model components, under the LDM, each component $H_j, \, j=1,\dots,p,$ gives rise to a univariate warping problem. To see this, consider for a fixed component $j$ the sample of univariate curves $S_j:=\{X_{ij}\}_{i=1}^n$. 
Using the normalized curves $X^*_{ij} = X_{ij}/||X_{ij}||_\infty$, estimation of $\gamma_j$ and $H_i$ for the $j^{th}$ component is a consequence of 
\begin{align}
X^*_{ij}(t) &= (\lambda\circ\Psi_j \circ H_i)(t), 
\label{eq_univ}
\end{align}
which coincides with a warping framework of type (\ref{eq.1}) with $\xi = \lambda\circ\Psi_j$, and $h_i = H_i$.
Replacing $X$ by $X^*$ in $(\ref{eq_univ})$ is necessary in order to eliminate the random amplitude factors $A_{ij}$. Since the random functions $G_{ij}$ are homeomorphisms, we have $||X_{ij}||_\infty = A_{ij} ||\lambda\circ G_{ij}||_\infty = A_{ij}$. Thus the normalized curves $X^*_{ij}(t) = (\lambda\circ\Psi_j\circ H_i)(t)$ do not depend on the  factors $A_{ij}$.


Applying an estimation method like pairwise warping for each of the subcollections $S_1,\dots,S_p$, results in $p$ estimates of the subject-level warping function, $\tilde{H}^{(1)}_i(t),\dots,\tilde{H}^{(p)}_i(t)$. Taking the mean of the resulting $p$ warping functions gives an estimate for the subject-specific warp, 
\begin{equation}
\hat{H}_i = p^{-1} \sum_{j=1}^p \tilde{H}^{(j)}_i, \quad i=1,\dots,n.
\end{equation}
\noindent For the  overall penalty parameter associated with the pairwise warping implementation we set 
\begin{equation}
\eta_1= \underset{1\leq j \leq p}{\max}~\eta_{1j}, 
\label{eq_eta1}
\end{equation}
 where $\eta_{1j}=10^{-4}\times  \{n^{-1} \sum_{i=1}^n 
\int_\mathcal{T} (X_{ij}(t)-\bar{X}_j(t))^2dt\}, ~j=1,\dots,p,$ is the default choice of the penalty parameter for each of the $p$ registrations, as per \cite{mull:08:6}. With  subject time warping estimators in hand, a plug-in estimate of $\gamma_j$ is obtained by averaging the component-aligned curves,  
\begin{equation}
\hat{\gamma}_j = n^{-1} \sum_{i = 1}^n (X_{ij}\circ \hat{H}_i^{-1})/||X_{ij}||_\infty,\quad\text{for}~j=1,\dots,p.
\end{equation}
\subsection{Global Alignment and Latent Curve Estimation}

A central idea in the estimation of the LDM is the fact that \textit{any} univariate curve $X_{ij}$ contains information about the latent curve, regardless of which component $j$ is considered.  This motivates a perspective in which we temporarily ignore the multivariate structure of the data and expand our scope to the full collection of curves,  $S=\cup_{j=1}^p S_j$. For each subject $i$, select one of its component curves at random as a representative. Call this representative curve $Z_i$ and denote its normalized counterpart by $Z^*_i$. Selecting one of the components at random ensures that we have $P(Z_i = X_{ij}) = 1/p$ for all $i=1,\dots,n,~j=1,\dots,p$. The collection of curves $\{Z_i,~i=1,\dots,n\}$ can be thought of as realizations of $\lambda$ subject to some random distortion $D_i$, where $D_i=G_{ij}$ if the $j^{th}$ component curve is selected. Define $I_{ij}$ as the event that the curve $Z_i$ comes from the collection of $j^{th}$ component curves, $S_j$. Conditional on the event $I_{ij}$ (which happens with probability $1/p$ for all $i=1,\dots,n$), it follows that $D_{i} = G_{ij}= \Psi_k
\circ H_i$.  Then, on average there is no net warping from the latent curve, as 
\begin{equation}
E[D^{-1}_{i}] = E\{E[D^{-1}_{i}|I_{ij}]\} = \sum_{j=1}^p E[H^{-1}_i\circ \Psi^{-1}_k] P(I_{ij}) =  p^{-1}\sum_{j=1}^p \Psi^{-1}_j = id.
\end{equation}
This observation motivates the warping problem 
\begin{equation}
Z^*_i = \lambda \circ D_i, \quad\text{for $i=1,\dots,n$}.
\end{equation}

The critical implication of this relation is that if we expand our scope to the full collection $S$ and apply a traditional method like pairwise warping to obtain $\hat{D}_{i}$ for all $i=1,\dots,n$, the latent curve can be estimated by averaging the globally-aligned curves, 
\begin{equation}
\hat{\lambda} = n^{-1} \sum_{i = 1}^{n} (Z_{i}\circ \hat{D}_{i}^{-1})/||Z_{i}||_\infty.
\end{equation}
\noindent The estimators of the component deformations are motivated by recalling that
\begin{align*}
{\gamma}_j = \lambda \circ \Psi_j , \quad  j=1,\dots.,p.
\end{align*}

\noindent Using a spline representation (see Section A of Appendix), we write 
\begin{equation}
\Psi_j(t) = \theta^T \alpha(t)
\label{eq:psispline}
\end{equation}
and estimate the component warps by solving the penalized minimization problem,
\begin{align}
\begin{split}
\tilde{\theta}_{\Psi_{j}} &= \underset{\theta\in\Theta}{\argmin} ~\mathscr{C}_{\eta_2}(\theta; \hat{\gamma}_j, \hat{\lambda}),\\
\mathscr{C}_{\eta_2}(\theta; \hat{\gamma}_j, \hat{\lambda})&=  \int_\mathcal{T} d^2\left(\hat{\gamma}_j,\hat{\lambda}(\theta^T\alpha(t))\right) dt+\eta_2\int_\mathcal{T}(\theta^T\alpha(t)-t)^2dt,
\end{split}
\label{eq:objfun2}
\end{align}
\noindent with 
$\eta_{2}=10^{-4}\times  \{p^{-1} \sum_{j=1}^p
\int_\mathcal{T} (\hat{\gamma}_j(t)-\hat{\lambda}(t))^2dt\}$ as the default choice of  penalty parameter in line with \cite{mull:08:6}. Finally, we obtain the component warps as 
\begin{equation}
\hat{\Psi}_j(t) = \tilde{\theta}_{\Psi_j}^T \alpha(t).
\end{equation}

\subsection{Measurement Error and Curve Reconstruction}
Note that under the assumption of fully observed curves without measurement error, the amplitude factors $A_{ij}=||X_{ij}||_\infty$ are known. Often in practice, this is not realistic, and the factors must be estimated by, e.g., $\hat{A}_{ij}=||\tilde{X}_{ij}||_\infty$ where $\tilde{X}$ denotes a smoothing estimate of a function $X$ that is observed with noise, as described in the following section. We note that these smoothing methods introduce a finite bias on the amplitude factors, but as the number of time points in the observation grid goes to infinity, our proposed estimate is asymptotically unbiased as shown in Theorem 1f. of Section \ref{sec:theory}. We refer to the Appendix for a detailed discussion of applying smoothing methods with the LDM.

After the smoothing step, estimates are obtained by substituting the smoothed curves in for $X_{ij}$ and implementing the procedure described in Sections 3.2 and 3.3. Once all model components are estimated,
plug-in estimates of the composite distortion functions and marginal and subject-level component deformation functions are an immediate consequence, 
\begin{align}
\hat{G}_{ij} &= \hat{\Psi}_j\circ \hat{H}_i, \label{eq:mixedwarp_est}\\
\hat{T}_{jk} &= \hat{\Psi}_j^{-1}\circ\hat{\Psi}_k,\label{eq:xct_est}\\
\hat{T}^{(i)}_{jk} &= \hat{G}_{ij}^{-1}\circ\hat{G}_{ik}, \quad \quad i=1,\dots,n, \quad j,k=1,\dots,p. \label{eq:subjxct_est}
\end{align}
\noindent Additionally, fitted curves based on the LDM can be obtained as
\begin{align}
\begin{split}
\hat{X}_{ij}(t) &= \hat{A}_{ij} (\hat{\lambda}\circ\hat{G}_{ij})(t)\\
&=\hat{A}_{ij} (\hat{\lambda}\circ\hat{\Psi}_{j}\circ \hat{H}_i)(t), \quad \quad i=1,\dots,n, \quad j,k=1,\dots,p.
\end{split}
\label{eq:recon}
\end{align}
These fits can be viewed through the lens of dimension reduction as their calculation require only $n+p+1$ estimated functions as opposed to $np$ curves in the original data.  This constitutes a novel representation for multivariate functional data that is distinct from the common functional principal component representations.
\vspace{-1cm}
\section{Data Applications}
\label{sec:data}
\subsection{Z\"urich Growth Study}

From 1954 to 1978, a longitudinal study on human growth and development was conducted at the University Children’s Hospital in Z\"urich. The sitting heights, arm lengths, and leg lengths of a cohort of children were measured on a dense time grid and these data can be viewed as densely sampled multivariate functional data. We focus on the timing of pubertal growth spurts, which usually occur between ages 9 and 18. It is standard in the growth curve literature to examine the derivatives of the growth curves, i.e. the growth velocities, instead of the curves themselves \citep{gass:84}. The velocities have a peak during puberty, with the crest location representing the age when an individual is growing fastest. 

\begin{figure}[!t]
	\centering
	\includegraphics[width=\linewidth]{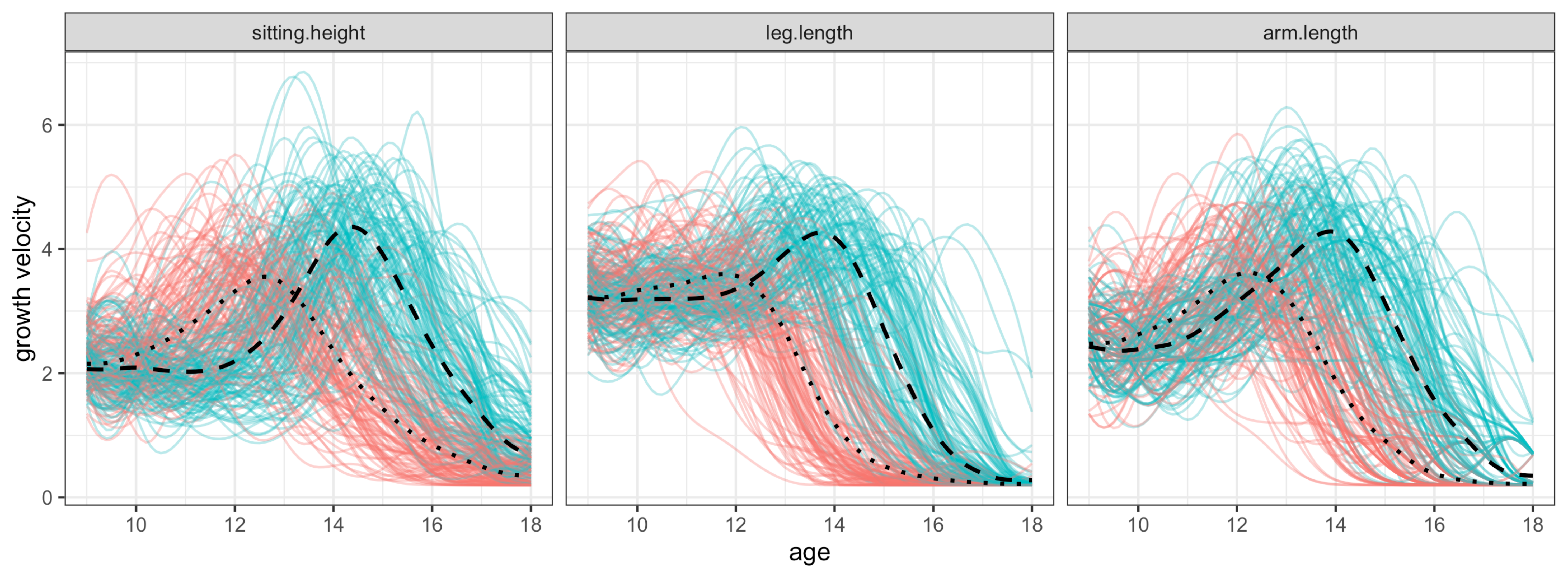}
	\caption{Growth velocities (in cm/year) during puberty for boys (blue) and girls (red). Scaled component tempo functions are marked for boys and girls with dashed and dotted lines, respectively. }
	\label{fig:growthvel}
\end{figure}

The timings and curvatures of these peaks are critical in informing growth patterns. In a first step, we estimated these growth velocities by local linear smoothing (Fig. \ref{fig:growthvel}).  It is well known that there is a difference in the pubertal growth patterns of boys and girls. This distinction is clear from just a simple inspection of the growth velocities in Figure 1. It is then of scientific interest, with practical implications for  auxologists, pediatricians and medical practitioners, to further study and quantify the differential between the onset of puberty for boys and girls, differentiated by different body parts. 

For the Z\"urich Longitudinal Growth Study, the biological clocks accelerate and deviate from clock time rapidly between the ages of 9 and 12 for girls and  between the ages of 12 and 15 for boys  (represented by the black dashed line on the diagonal). Component tempos for boys and girls are a simple way to summarize these differences (Fig. \ref{fig:growthvel}, dashed and dotted lines, respectively), as they serve as the structural means of the timing functions. 

\begin{figure}[!t]
	\centering
	\includegraphics[width=\linewidth]{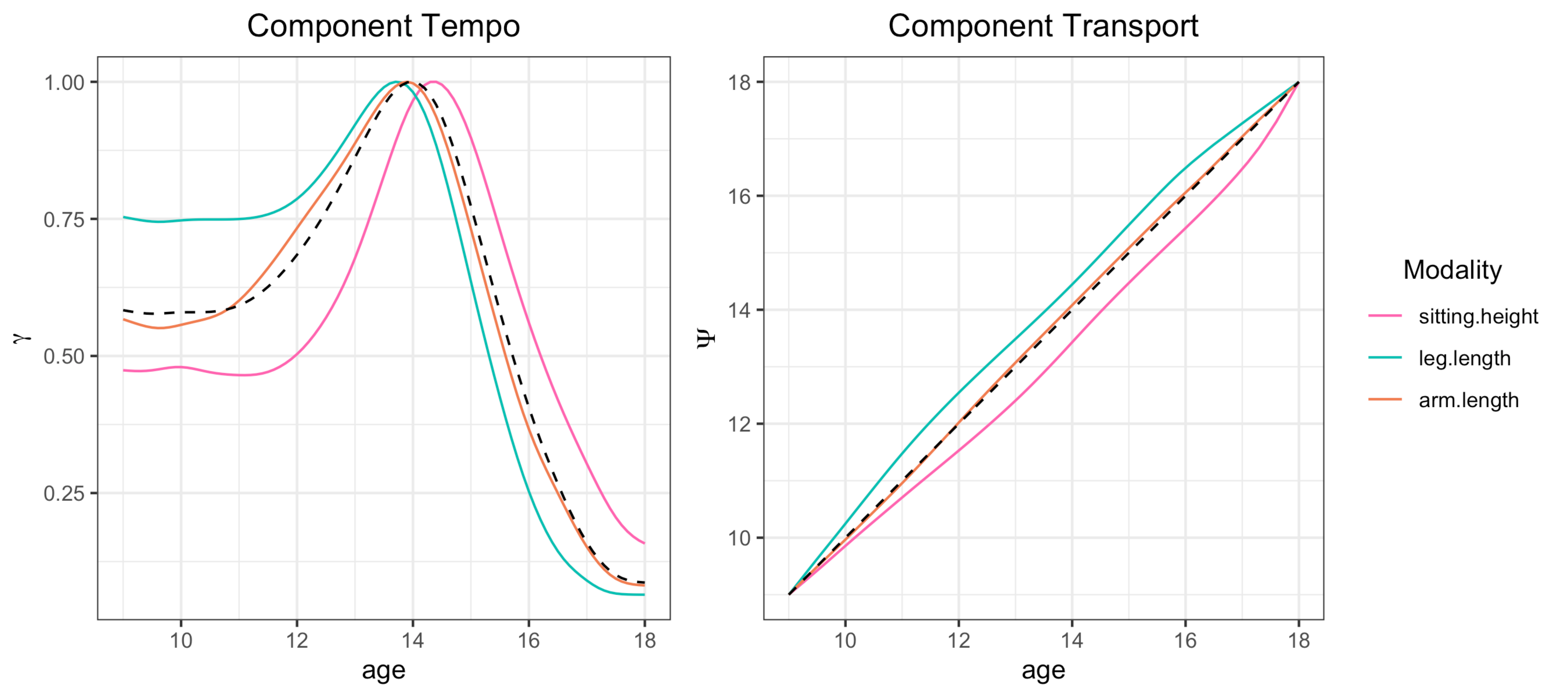}
	
	\caption{Component tempos $\gamma$ (left) and deformations $\Psi$ (right) for growth modalities. The dashed line represents the tempo and deformation for the latent tempo, $\lambda$. }
	\label{fig:bary}
\end{figure}
Considering the joint time dynamics of the $p=3$ modalities, we restrict our analysis to the boys for the sake of brevity. A natural place to start when comparing growth patterns is the component tempos, which are displayed for each modality in the left panel of Fig.~\ref{fig:bary}. The dynamics of joint development emerges when examining the order of peaks across modalities. Leg length is first, followed by arm length, while sitting height lags behind. The tempos have similar slopes during puberty, though leg length has the most gradual spurt and sitting height the sharpest, perhaps because its lagged onset results in a smaller window between the onset of its growth spurt and the maturation date of 18 years. While it is  possible for an individual to experience some minor growth past the age of 18,  in the Z\"urich study such cases were rare and so this complication was ignored. The component deformations displayed in Fig.~\ref{fig:bary} (right) further illustrate the nature of each body part's tempo relative to baseline.  Remarkably, the tempo of arm length is nearly identical to the latent curve. This suggests that the arm can be used a representative modality which mirrors a child's overall development. 



\begin{figure}[!t]
	\centering
	\includegraphics[width=\linewidth]{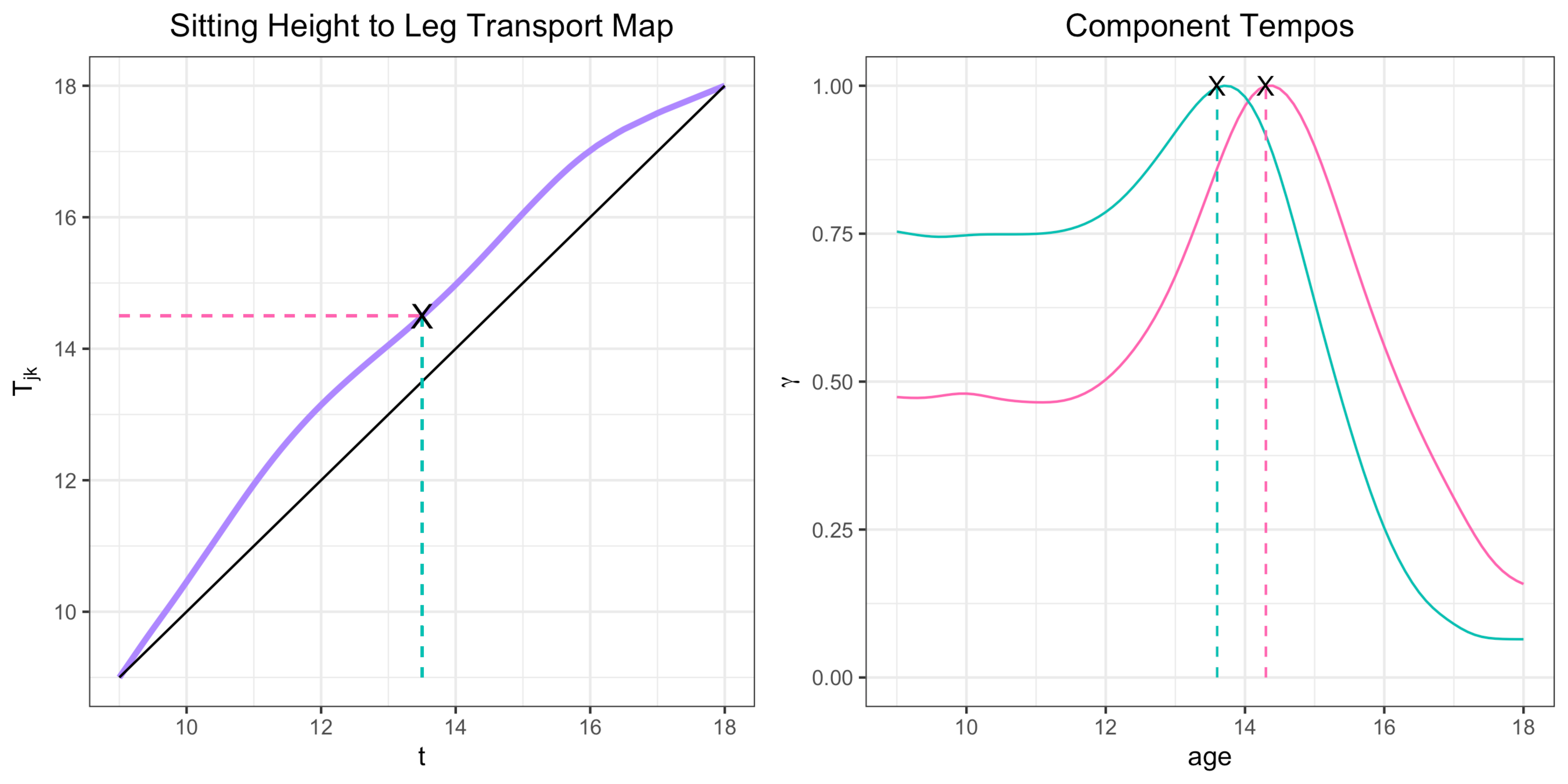}
	
	\caption{The cross-component deformation map $T_{12}$ which expresses the sitting height's timing patterns relative to the leg length's as a baseline. The peak of pubertal growth rate for the leg occurs at approximately age $13.5$, while the maximum growth velocity for sitting height growth occurs at approximately $T_{12}(13.5) \approx 14.5$ years old.}
	\label{fig:sit_to_leg}
\end{figure}

We also can interpret the cross-component deformations, $\hat{T}_{jk},~j,k\in\{1,\dots,p\}$, estimated as per (\ref{eq:xct_est}). The magnitude of the XCD map's deviation from the identity shows how dissimilar two components are. For example, sitting height and leg length are the most distinct modalities of growth among those considered here, and their XCD map exhibits the most pronounced departure from the identity. An intuitive interpretation of the map is that $T_{jk}$ expresses the $k^{th}$ component's timing patterns relative to the $j^{th}$ component's as a baseline. For example, when the leg tempo is at time $t=13.5$, the comparable time point for the sitting height tempo is approximately at $T_{jk}(13.5) \approx 14.5$,  as illustrated in Fig.~$\ref{fig:sit_to_leg}$. 



\subsection{Air Pollutants in Sacramento, CA}

The study of air pollutants has been a topic of interest for atmospheric scientists and environmentalists alike for several decades. In particular, increased ground-level ozone (O$_3$) concentrations have been shown to have harmful effects on human health.  Unlike many air pollutants, surface ozone is not directly emitted by sources of air pollution (e.g. road traffic); it is formed as a result of interactions between nitrogen oxides and volatile organic compounds in the presence of sunlight \citep{abdul:01}. Because of this interaction, compounds such as nitrogen dioxide are known and important precursors of increased ozone concentrations \citep{tu:07}.

\begin{figure}[!t]
	\centering
	\includegraphics[width=\linewidth]{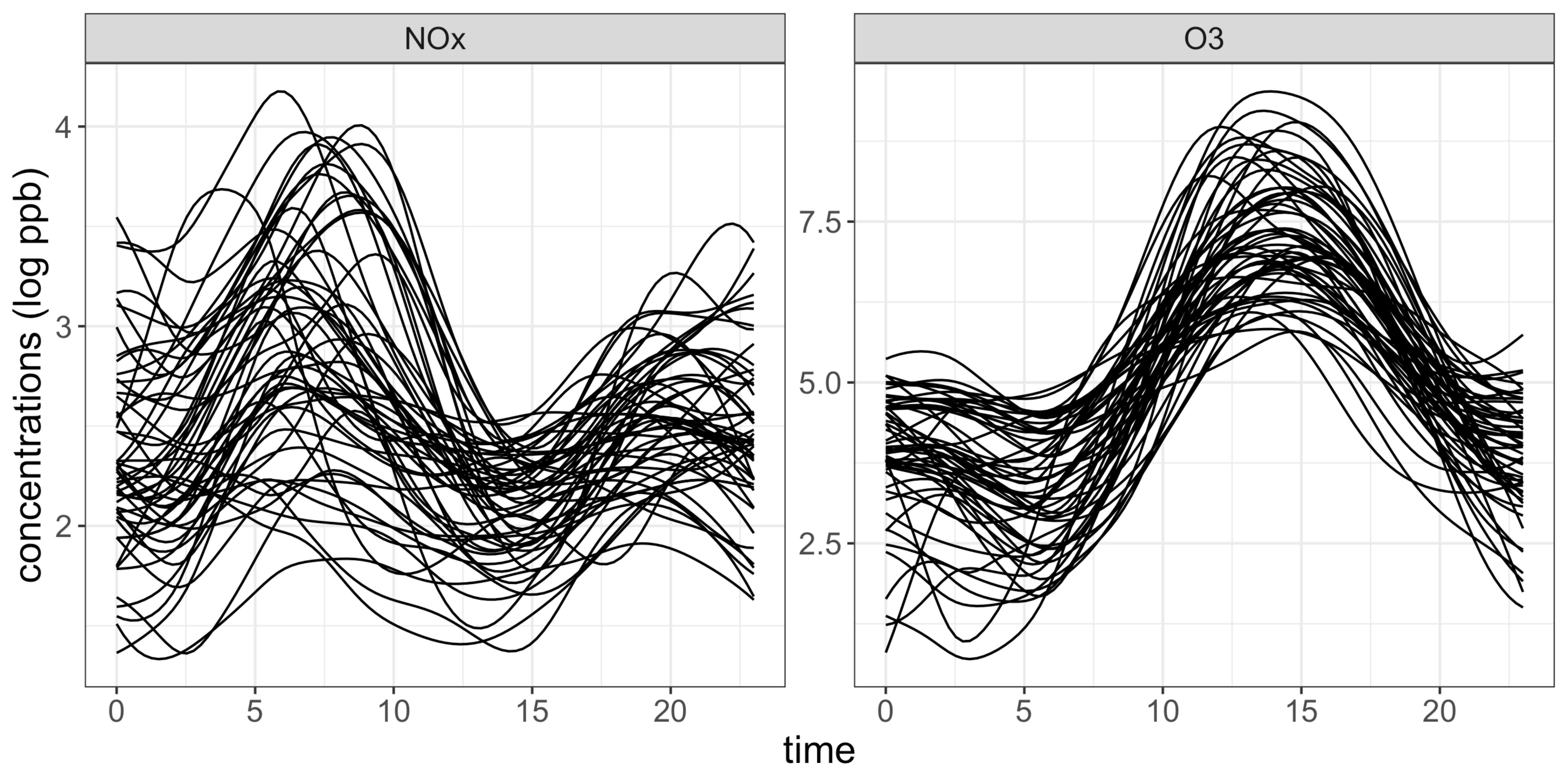}
	\caption{24-hour trajectories of NO$_x$ (left) and ozone (right),  concentrations in parts per billion (ppb) on a log scale.}
	\label{fig:air_trajectory}
\end{figure}

The California Environmental Protection Agency has monitored hourly air pollutant concentrations at several station locations since the 1980s. Here we consider the sample of weekday trajectories of ozone (O$_3$), and nitrogen oxides (NO$_x$) concentrations during the summer of 2005 in Sacramento (Fig. \ref{fig:air_trajectory}). Smooth trajectories were obtained from raw data using local linear weighted least squares. \cite{gerv:15} has previously investigated a similar dataset in the context of warped functional regression, where the primary aim was to model phase variation explicitly in order to relate the timing of peak concentrations of NO$_x$ to those of O$_3$.

The chemistry of the compounds as well as a visual inspection of the curves suggests that the are two distinct classes of pollutants. NO$_x$ concentrations tend to peak around 8 a.m., reflecting standard morning commute hours and the impact of traffic emissions  on air quality. On the other hand, ozone levels peak around 2 to 3 p.m., indicating that the synthesis mechanism induces a lag of up to approximately 6 hours.

It is then of interest to study whether  meteorological factors might affect the rate of ozone synthesis. Individual component deformations combined with Fr\'echet regression for distributions provide a natural framework for this  \citep{pete:19}. Subject-specific deformations from NO$_x$ concentrations to ozone concentrations, $T^{(i)}_{NO_x\to O_3}$, were calculated as per (\ref{eq:subjxct_est}) for each day. Global Fr\'echet regression was then applied through fitting the model
\begin{align}
\begin{split}
\hat{m}_\oplus(x) &= \underset{T\in\mathcal{W}}{\argmin}~M_n(T,x),\\
M_n(T,x) &= n^{-1} \sum_{i=1}^n q_{in}d^2_W(T_i,T),
\end{split}
\end{align}
where $m_\oplus$ denotes the conditional Fr\'echet mean of the deformation given the covariate $x$, the wind speed recorded a given day. Here, $d_W$ is  the $2-$Wasserstein distance \citep{vill:03} and the weights $q_{in}$ are derived from global linear regression and defined as $q_{in}=1+(x_i-\bar{x})(x-\bar{x})/\hat{s}^2_x$ \citep{pete:19}, where $\bar{x}$ and $\hat{s}^2_x$ represent the sample mean and variance of the observed wind speeds, respectively. The model was fit using the \texttt{R} package \texttt{frechet}, observing that the deformation functions can be represented as distribution functions \citep{frechet}. 

\begin{figure}[!t]
	\centering
	\includegraphics[width=\linewidth]{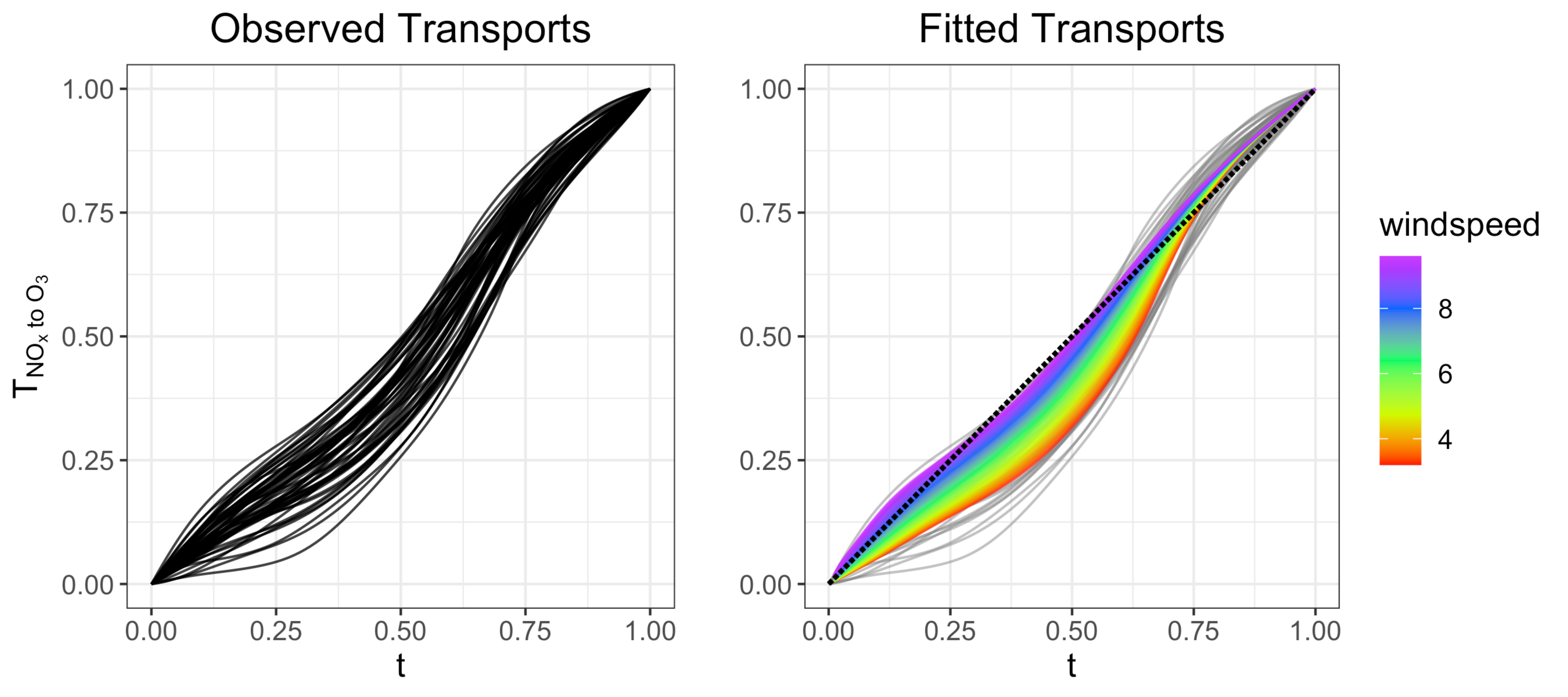}
	\caption{Fr\'{e}chet regression of NO$_x$-to-O$_3$ cross-component deformations onto daily max windspeeds in knots. Windier days correspond to more linear deformation functions, which suggests  O$_3$ synthesis more closely follow NO$_x$ emission. Less windy days are associated with more pronounced lags between the pollutants. }
	\label{fig:freg}
\end{figure}

Figure $\ref{fig:freg}$ displays the observed deformations and the fits obtained from Fr\'echet regression using windspeed as a predictor. The rainbow gradient corresponds to windspeeds ranging from 3 to 10 knots and their associated fitted deformations are overlaid the original data. The regression fits suggest that days with lower windspeeds correspond with deformations which are further from the diagonal, indicating an exaggerated lag between peak concentrations of NO$_x$ and ozone. On the other hand, days with high wind speeds have fitted deformations very near the diagonal which suggests that windier settings accelerate the synthesis process. Intuitively this is a reasonable result in terms of the physical interpretation, as more wind will result in a higher rate of collisions of the particles, and thus quicker production of ozone after peak NO$_x$ emission. The Fr\'echet $R^2_\oplus$ value was $0.44$, which suggests that wind speed explains a considerable amount of variation in the observed deformations. 
\vspace*{-0.75cm}
\section{Theoretical Results}
\label{sec:theory}

Our results focus on convergence of the components of the LDM described in (4) as the number of curves $n$ and the number of observations per curve $m$ tend to infinity. We require the following assumptions on (L) the components of the Latent Deformation Model and (S) the smoothing methodology in the presence of discretely observed curves:

\begin{enumerate}[label=(L\arabic*), mode=unboxed]
	\setcounter{enumi}{0}
	
	\item The latent curve $\lambda(t) \in C^2(D)$ is a bounded function. For any non-degenerate interval $\mathcal{T}_0\subset \mathcal{T}, ~0<\int_{\mathcal{T}_0}\lambda'(t)^2dt < \infty$. 
	
	\item For $j=1,\dots,p$,  $\underset{1\leq i \leq n}{\sup} ~A_{ij} = \mathcal{O}_P(1)$ and $\underset{1\leq i \leq n}{\sup} ~A^{-1}_{ij} = \mathcal{O}_P(1)$. 
\end{enumerate}

Assumption (L1) bounds the latent curves and its derivatives and ensures there are no flat stretches and the 
uniqueness of the component estimates. (L2) ensures that the ranges of the random processes are bounded away from zero and infinity with high probability; this condition is needed for the uniform convergence of the smoothing estimate. 




\begin{enumerate}[label=(S\arabic*), mode=unboxed]
	\setcounter{enumi}{-1}
	
	\item The time points $t_1,\dots,t_m$, 
	depend on the sample size $n$, $m=m(n)$ and   constitute a dense regular design with smooth design density $f$ with $\underset{t \in \mathcal{T}}{\inf }f(t)>0$ that generates the time points according to $t_s = F^{-1}(\frac{s-1}{m-1}), ~s=1,\dots,m,$ where $F^{-1}$ denotes the quantile function associated with $f$. The second derivative $f''$ is bounded, $\underset{t \in \mathcal{T}^\circ}{\sup}|f''(t)|<\infty$.	
	
	\item The kernel function $K$ is a probability density function with support $[-1,1]$, symmetric around zero, and uniformly continuous on its support,  with $\int_{-1}^1 K^2(u)du<\infty$.

	\item For each $j=1,\dots,p$, the sequences $m = m(n)$ and  $b = b(n)$ satisfy (1) $0<b< \infty$, and (2) $m\to\infty$, $b\to0$, and $mb^2(\log b)^{-1}\to\infty$ as $n\to\infty$.
	
\end{enumerate}

\noindent  These assumptions guarantee the consistent estimation of $n$ curves simultaneously, as shown in the following Proposition. We observe that (S2) is for example satisfied if the bandwidth sequence is chosen such that $b=b(n) \sim m(n)^{-1/6}$. 

\begin{proposition}
	Under assumptions (S0$-$S2), if $E||X^{(\nu)}(t)||_\infty^2<\infty, ~\nu=0,1,2$,  we have the uniform convergence
	\begin{equation}
	\underset{t \in \mathcal{T}}{\sup}|\tilde{X}_{ij}(t)-X_{ij}(t)| = \mathcal{O}_P(m^{-1/3}).
	\end{equation}
	The rate also extends to the standardized versions $X_{ij}^* = X_{ij}/||X_{ij}||_\infty$, 
	\begin{equation}
	\underset{t \in \mathcal{T}}{\sup}\left|\frac{\tilde{X}_{ij}(t)}{||\tilde{X}_{ij}||_\infty}-\frac{X_{ij}(t)}{||X_{ij}||_\infty}\right| = \mathcal{O}_P(m^{-1/3}).
	\end{equation}
\end{proposition}

This result agrees with the existing results in the literature, in that it is a special case of a general result for metric-space valued functional data  (see \cite{mull:20}),  now here in the case of real-valued functions. The estimators of the latent curve and component deformations involve averages of the smoothing estimates over the sample of curves as $n\to\infty$. The corresponding rates of convergence will thus rely on the uniform summability of the difference between the smoothed and true curves over $n$ and we then have a uniform rate of $\tau_m=m^{-(1-\delta)/3}$ for an arbitrarily small $\delta>0$ in lieu of the above rate $m^{-1/3}$; see Lemma 1 in the Appendix. 
The proposed estimators also rely on the mechanics of the pairwise warping methods, whose convergence properties have been established in a general form in \cite{mull:08:6} and \cite{mull:20}. Lemma 2 in the Appendix states these rates in the specific framework of the Latent Deformation Model. 
We are now in a position  to state our main result, which establishes rates of convergence for the estimators of the components of the Latent Deformation Model as follows.  

\begin{theorem}
	Under assumptions  (L1), (L2), and (S0$-$S2), with $\tau_m=m^{-(1-\delta)/3}$ for an arbitrarily small $\delta>0$ and penalty parameters as described in (\ref{eq_eta1}) and  (\ref{eq:objfun2}),  we have for all $i=1,\dots,n,~j=1,\dots,p$, 
	
	a. $\underset{t \in \mathcal{T}}{\sup}|\hat{H}_{i}(t) - H(t)| = \mathcal{O}_P(n^{-1/2}) + \mathcal{O}_P(\tau_m^{1/2})+ \mathcal{O}(\eta_1^{1/2})$,
	
	b. $\underset{t \in \mathcal{T}}{\sup}|\hat{\gamma}_j(t) - \gamma_j(t)| = \mathcal{O}_P(n^{-1/2}) +  \mathcal{O}_P(\tau_m^{1/2})+ \mathcal{O}(\eta_1^{1/2}) $
	
	c. $\underset{t \in \mathcal{T}}{\sup}|\hat{\lambda}(t) - \lambda(t)| = \mathcal{O}_P(n^{-1/2})+  \mathcal{O}_P(\tau_m^{1/2})+ \mathcal{O}(\eta_1^{1/2})$, 
	
	d. $\underset{t \in \mathcal{T}}{\sup}|\hat{\Psi}_j(t) - \Psi_j(t)| = \mathcal{O}_P(n^{-1/2}) +  \mathcal{O}_P(\tau_m^{1/2})+ \mathcal{O}(\max(\eta_1,\eta_2)^{1/2})$,  
	
	e. $\underset{t \in \mathcal{T}}{\sup}|\hat{G}_{ij}(t) - G_{ij}(t)| = \mathcal{O}_P(n^{-1/2}) +  \mathcal{O}_P(\tau_m^{1/2})+ \mathcal{O}(\max(\eta_1,\eta_2)^{1/2})$, and 
	
	f. $|\hat{A}_{ij} - A_{ij}| = \mathcal{O}_P(m^{-1/6}).$
\end{theorem}

The three terms in the rates correspond, in order, to (1) the parametric rate achieved through the standard central limit theorem, (2) the smoothing rate which is dependent on the number of observations per curve $m$, and (3) a rate due to the well-known bias introduced by the penalty parameters used in the regularization steps. Additionally, if we suppose that $m$ is bounded below by a multiple of $n^{3(1-\delta)^{-1}}$, then the rates corresponding to the smoothing steps are bounded above by $n^{-1/2}$. If we take the penalty parameters to be $\eta_1\sim\eta_2=\mathcal{O}(n^{-1})$, a $n^{-1/2}$ rate of convergence can be achieved for each of the estimators in Theorem 1~$a.$-$e.$ Otherwise if $m\sim n^{\Delta(1-\delta)^{-1}}$, for any $\Delta<3$,  the convergence is limited by the smoothing step and achieves the rate of $n^{-\Delta/6}$. 



\begin{corollary1}
	Suppose the penalty parameters $\eta_1\sim\eta_2=\mathcal{O}(n^{-1})$. If the random trajectories are fully observed without error or the trajectories are recorded with at least a multiple of $m\sim n^{\Delta(1-\delta)^{-1}}$  observations per curve, with $\Delta>3$, then under the assumptions of Theorem 1, we have for~all~$i=1,\dots,n,~j=1,\dots,p$, 
	
	a. $\underset{t \in \mathcal{T}}{\sup}|\hat{H}_{i}(t) - H(t)| = \mathcal{O}_P(n^{-1/2})$,
	
	b. $\underset{t \in \mathcal{T}}{\sup}|\hat{\gamma}_j(t) - \gamma_j(t)| = \mathcal{O}_P(n^{-1/2}) $
	
	c. $\underset{t \in \mathcal{T}}{\sup}|\hat{\lambda}(t) - \lambda(t)| = \mathcal{O}_P(n^{-1/2})$, 
	
	d. $\underset{t \in \mathcal{T}}{\sup}|\hat{\Psi}_j(t) - \Psi_j(t)| = \mathcal{O}_P(n^{-1/2})$, 
	
	e. $\underset{t \in \mathcal{T}}{\sup}|\hat{G}_{ij}(t) - G_{ij}(t)| = \mathcal{O}_P(n^{-1/2})$, and 
	
	f. $|\hat{A}_{ij} - A_{ij}| = \mathcal{O}_P(n^{-1/2}).$
	
\end{corollary1}

The asymptotic results for the cross-component deformations then follow immediately from the rates established in Theorem 1. 

\begin{theorem}
	Under assumptions of Theorem 1 for~~$i=1,\dots,n,~1\leq j,k\leq p$,
	
	a.
	$\underset{t \in \mathcal{T}}{\sup}|\hat{T}_{jk}(t)- T_{jk}(t)| = \mathcal{O}_P(n^{-1/2}) +  \mathcal{O}_P(\tau_m^{1/2})+ \mathcal{O}( \max(\eta_1,\eta_2)^{1/2}),$\quad\text{and}
	
	b. $\underset{t \in \mathcal{T}}{\sup}|\hat{T}^{(i)}_{jk}(t)- T^{(i)}_{jk}(t)| = \mathcal{O}_P(n^{-1/2}) +  \mathcal{O}_P(\tau_m^{1/2})+ \mathcal{O}( \max(\eta_1,\eta_2)^{1/2}).$
\end{theorem}

A similar corollary for cross-component deformations follows in the case of fully observed curves or dense enough designs. 

{\begin{corollary1}
	Suppose the penalty parameters $\eta_1\sim\eta_2=\mathcal{O}(n^{-1})$. If the random trajectories are fully observed without error or are recorded with at least a multiple of $m\sim n^{\Delta(1-\delta)^{-1}}$  observations per curve, with $\Delta>3$, then under the assumptions of Theorem 1, we have for~~$i=1,\dots,n,~1\leq j,k\leq p$,
	
	a.	$\underset{t \in \mathcal{T}}{\sup}|\hat{T}_{jk}(t)- T_{jk}(t)| = \mathcal{O}_P(n^{-1/2}),$\quad\text{and}
	
	b. $\underset{t \in \mathcal{T}}{\sup}|\hat{T}^{(i)}_{jk}(t)-T^{(i)}_{jk}(t)| = \mathcal{O}_P(n^{-1/2}).$
	
	\end{corollary1}
}

Corollaries 1 and 2 suggest that, on dense enough measurement schedules, parametric rates of convergence are achievable for the components of  the LDM. 

\begin{remark1}
	For any cycle of components indexed by the sequence, $$\pi_1\to \pi_2\to \pi_3\to\dots \to \pi_L\to \pi_1,$$ with arbitrary length $L$ and $\pi_1,\dots,\pi_L\in \{1,\dots, p\}$, their respective cross-component deformations satisfy 
	$$T_{\pi_1\pi_2} \circ T_{\pi_2\pi_3} \circ \dots \circ T_{\pi_L\pi_1} = id.$$
\end{remark1}
\noindent This result ensures that the system of cross-componentdeformations maps prevents inconsistencies within itself. For example, if for three components $A$, $B$, and $C$, the pairwise deformations $T_{AB}$ and $T_{BC}$ suggest that Component $A$ tends to precede Component $B$ which itself tends to precede Component $C$, this implies that the deformations $T_{AC}$ must indicate that Component $A$ tends to precede Component $C$. Furthermore, mapping a component tempo through other components and then back to itself will result in the original component tempo, unchanged.
Next we consider the convergence rates of reconstructed curves as per (\ref{eq:recon}), putting all model components together. 
\begin{theorem}
	Under assumptions of Theorem 1 for~~$i=1,\dots,n,~j=1,\dots ,p$,
	$$\underset{t \in \mathcal{T}}{\sup}|\hat{X}_{ij}(t)-X_{ij}(t)|= \mathcal{O}_P(n^{-1/2}) +  \mathcal{O}_P(\tau_m^{1/2})+ \mathcal{O}( \max(\eta_1,\eta_2)^{1/2}).$$
\end{theorem}
Again a parametric rate is achievable on dense enough designs.

\begin{corollary1}
	Suppose the penalty parameters $\eta_1\sim\eta_2=\mathcal{O}(n^{-1})$. If the random trajectories are fully observed without error or the trajectories are recorded with at least a multiple of $m\sim n^{\Delta(1-\delta)^{-1}}$  observations per curve, with $\Delta>3$, then under the assumptions of Theorem 1, we have for~$i=1,\dots,n,~j=1,\dots,p$,
	$$\underset{t \in \mathcal{T}}{\sup}|\hat{X}_{ij}(t)-X_{ij}(t)|= \mathcal{O}_P(n^{-1/2}).$$
\end{corollary1}

\section{Concluding Remarks}

The Latent Deformation Model (LDM) provides a novel decomposition for a large class of practically relevant multivariate functional data by quantifying their inter-component time dynamics. A separability assumption that makes it possible to factor overall time warping into component-specific and subject-specific time warping components is crucial.  The ensuing simple representation for multivariate functional data includes 
two fixed effect terms (the latent curve and a collection of component-level warping functions) and two random effect terms (a random amplitude vector and a collection of subject-level warping functions). This representation requires the estimation of only one random warping function and amplitude vector per subject, in addition to $p+1$ deterministic functions overall.

In some cases these components may be reduced even further. For example, when subject-level warping is negligible or part of a pre-processing step, a special case of the model arises in which time dynamics are fully characterized by the $p+1$ fixed effect curves and one random scalar per component. Alternatively, if subject-level time warping is present but further dimension reduction is desired, transformation of warps by the LQD transform \citep{mull:16:1} or other means  (see, e.g. \cite{happ:19}) will permit a Karhunen-Lo\`eve expansion in $\mathcal{L}^2-$space. Applying the LDM and truncating this expansion at an appropriate number of eigenfunctions, say $K_0$, creates a representation of multivariate functional data using only $p+K_0$ random scalars, as opposed to  a standard FPCA representation which requires $p\times K_0$ variables. 

A limitation of this framework  is the fact that slight deviations from a common latent curve will always occur in practice. An implicit assumption in applying the LDM is that the magnitude of nuisance peaks is negligible in comparison to the dominant features of the latent curve. Simulations which examine the robustness of component estimates in the presence of model misspecification or more pronounced nuisance peaks are  in the supplement. 

The LDM serves both as an extension of existing univariate functional warping methods, as well as a stepping stone for many new potential models for multivariate functional data analysis and registration. Future directions of note include harnessing cross-component deformation maps for imputating components in partially observed multivariate functional data, or relaxing structural assumptions to allow for more flexible functional relationships between different latent curves for distinct subsets of components; e.g. allowing for multiple latent curves, $\lambda_1(t), \lambda_2(t)$, with $\lambda_1(t) = g(\lambda_2(t))$ for some function $g$. Spatiotemporal applications are also promising for the LDM, in which the vector components are indexed by location. Then component warping functions may reveal time trends across geographic regions. 

\vspace{-0.75cm}
\section*{Acknowledgments}

We wish to thank two referees for very useful suggestions. This research was supported in part by NSF grant DMS-2014626.

\vspace{-0.75cm}
\section*{Data Availability Statement}
The air pollutant data used in the application section are publicly available on the California Air Resource Board's website: https://www.arb.ca.gov/adam. The growth curve data are proprietary to the Z\"urich University Children's Hospital and therefore not shared.
 \vspace{-0.75cm}
 
 \newpage
\bibliographystyle{asa}
\bibliography{syncfunc}


\label{lastpage}

\end{document}